\documentstyle[aps]{revtex}

\preprint {UTF/98-05-416} 

\title{Quark contribution to the nucleon polarizabilities\\
and three-body forces}

\author{Elvira Biasioli$\,^{1}$, Marco Traini$\,^{1,}$
\thanks{{\it E-mail address:} traini@science.unitn.it} 
and Renzo Leonardi$\,^{1,2}$}
\address{
$^1$ Dipartimento di Fisica, Universit\`a degli Studi di Trento, 
I-38050 POVO
(Trento), Italy\\ {\rm and} Istituto Nazionale di Fisica Nucleare, 
G.C. Trento\\
$^2$ ECT$\star$, European Centre for Theoretical Studies in Nuclear Physics 
and Related Areas,\\
Villa Tambosi, I-38050 VILLAZZANO (Trento), Italy}

\begin{document}

\maketitle

\begin{abstract}
We study the response of the nucleon, as a system of three bound (constituent) 
non relativistic quarks, to external (quasi static) electric and magnetic fields.
The approach, based on a sum rule technique, is applied to a large class of two and 
three-body interquark potentials. Lower and upper bounds to the electric 
polarizability and para-magnetic susceptibility are explicitly calculated within 
a large variety of constituent models and their values related to the features of
the interquark interaction picture.  The r\^ole of three-body forces is investigated
in details as well as the effects of $SU(6)$ breaking terms in the potential model.
Our results can be used to extract the mesonic contributions to the static polarizabilty 
and susceptibility. The quark degrees of freedom give a quite sizeable contributions to both
and the meson cloud accounts roughly for $30\%$ and $60\%$ of the electric proton and 
neutron polarizability respectively. The quark contribution to the paramagnetic susceptibility
is even higher and the mesonic effects are rather uncertain.
\end{abstract}

\section{Introduction}
\label{intro}

The electric and magnetic polarizabilities, labeled $\alpha_E$ and $\beta_M$
respectively, are fundamental observables to understand
the intrisic structure of the nucleons. They caracterize the ability of the 
constituents to rearrange in response to (quasi) static external electric 
and magnetic fields and are as fundamental as other parameters like charge 
radii and magnetic moments.

For the proton they can be measured by means of Compton scattering experiments 
since the Thomson scattering amplitude $T_{fi} = - \hat \epsilon \cdot \hat \epsilon'
\,e^2/M_N$ is modified, for intermediate values of the photon energies 
($50$ MeV $\lesssim \omega \lesssim 100$ MeV), by first order corrections due to the
nucleon polarizabilities yielding to the low-energy Compton scattering cross section
\cite{comptonCS}
\begin{eqnarray}
{d\sigma \over d\Omega}(\omega,\theta) & = & {d\sigma^B \over d\Omega}(\omega,\theta) - 
{e^2\over M_N} \left({\omega'\over\omega}\right) \left({\omega' \omega}\right) \nonumber \\
& \times & \left[ {\bar \alpha_E^{\rm p} + \bar \beta_M^{\rm p} \over 2}\,(1+\cos\theta)^2 +
{\bar \alpha_E^{\rm p} - \bar \beta_M^{\rm p} \over 2}\,(1-\cos\theta)^2 \right]\,\, ,
\label{CCS}
\end{eqnarray}
where ${d\sigma^B / d\Omega}$ is the Born cross section for a proton with an anomalous 
magnetic moment and no other structure (e.g. ref.\cite{drechsel97}), 
and $e^2$ is the fine structure constant.
$\bar \alpha_E$ and $\bar \beta_M$ are the so called dynamic (or Compton) polarizabilities
containing (within a non-relativistic approach) retardation effects as well as diamagnetic 
contributions \cite{corrections}. We will discuss such effects later on.

Eq.~(\ref{CCS}) shows that forward and backward cross sections are dominated by 
${\bar \alpha_E^{\rm p} + \bar \beta_E^{\rm p}}$ and ${\bar \alpha_E^{\rm p} - 
\bar \beta_M^{\rm p}}$ respectively, while 
experiments at $90^0$ are sensitive to $\bar \alpha_E^{\rm p}$ only and the polarizabilities are
obtained by measuring the deviations of the cross section from the Born values. In addition 
a dispersion sum rule \cite{baldinSR} constrains the sum of 
$\bar \alpha_E^{\rm p}$ and $\bar \beta_M^{\rm p}$:
\begin{equation}
\bar \alpha_E^{\rm p} + \bar \beta_M^{\rm p} = {1 \over 2 \pi^2}\,\int_{m_\pi}^\infty\,
{\sigma_{\rm tot}^{\rm p}(\omega)\,d\omega \over \omega^2} \approx 14.2\pm 0.5\,10^{-4}\,
{\rm fm}^3\,\,,
\label{baldsr}
\end{equation}
where $\sigma_{\rm tot}^{\rm p}(\omega)$ is the total proton photoabsorption cross 
section and the numerical value is obtained extrapolating tha available experimental 
data to infinite energy \cite{extrapol}. 

A global analysis of the existing experimental data has been recently 
performed \cite{mcgibbon95} yielding to
\begin{equation}
\bar \alpha_E^{\rm p} = (12.1\pm0.8\pm0.5)\,10^{-4}\,{\rm fm}^3;
\;\;\;\;\;\;\;\;\;
\bar \beta_M^{\rm p} = (2.1\pm0.8\pm0.5)\,10^{-4}\,{\rm fm}^3;
\label{ab-p}
\end{equation}

The neutron has no charge and Compton interference experiments to measure 
$\bar \alpha_E^{\rm n}$ 
are not possible\footnote{Compton scattering on the neutron can be realized by means of 
deuteron targets, e.g. ref.\cite{rose90}.}. Recent measurements (at ORNL) have been realized 
by means of a precise n-Pb scattering experiment \cite{ornl91}. The electric dipole moment, 
induced by the nuclear charge on the moving neutron, produces a secon order effect resulting 
in a $1/r^4$ interaction proportional to the neutron electric polarizability which can be 
measured looking at the energy dependence of the neutron-$^{208}$Pb scattering cross section.
The result is 
\begin{equation}
\bar \alpha_E^{\rm n} = (12.3\pm 1.5 \pm 2.0)\,10^{-4}\,{\rm fm}^3;
\;\;\;\;\;\;\;\;\;
\bar \beta_M^{\rm n} = (3.5\pm 1.6 \pm 2.0)\,10^{-4}\,{\rm fm}^3;
\label{ab-n}
\end{equation}
where $\bar \beta_M^{\rm n}$ is extracted by assuming $\bar \alpha_E^{\rm n} +
\bar \beta_M^{\rm n}=(15.8 \pm 0.5)\,10^{-4}$ fm$^3$ according to the sum 
rule (\ref{baldsr}), and the retardation effects are included following a suggestion
due to L'vov and Petrun'kin. The direct experimental result of the n-Pb scattering 
is $\alpha_E^{\rm n}$ rather than $\bar \alpha_E^{\rm n}$ since the measurement has a static 
character, and the value is $\alpha_E^{\rm n} = (12.0\pm 1.5 \pm 2.0)\,10^{-4}\,
{\rm fm}^3$ which is quite close the Compton polarizability because the corrections 
for the neutron are small.

For our study we will use non-relativistic approches and, in this case, the corrections 
assume a rather simple form. In fact the Compton polarizabilities can be
written (e.g. Friar in ref.\cite{comptonCS})
\begin{eqnarray}
\bar \alpha_E & = & {1\over 3 M}\,\langle 0|\sum_{i=1}^3e_i\,{{\bf r}_i'}^2|0\rangle 
+ 2 \sum_{n > 0} {|\langle n|D'_z|0\rangle|^2 \over E_n - E_0}=
\nonumber \\
& = & \Delta \alpha_E + \alpha_E
\label{alphae}
\end{eqnarray}
and
\begin{eqnarray}
\bar \beta_M & = & -{1\over 2 M}\,\langle 0|\left(\sum_{i=1}^3e_i\,{{\bf r}_i'}\right)^2|0\rangle 
-{1\over 6}\,\langle 0|\sum_{i=1}^3e_i^2\,{{\bf r}_i'}^2/m_i|0\rangle 
+ 2 \sum_{n > 0} {|\langle n|\mu_z|0\rangle|^2 \over E_n - E_0}
\nonumber \\
& = & \left.\left.\Delta \beta_M + \beta_M\right|_{\rm dia}+\beta_M\right|_{\rm para}\,\,.
\label{betam}
\end{eqnarray}

In Eq.~(\ref{alphae}) 
\begin{equation}
D'_z = \sum_{i=1}^3\,e_i z_i'
\label{DzN}
\end{equation}
is the nucleon electric dipole operator in the $z$-direction, i.e. in the direction of the 
external electric field. The coordinate of the charges $z'_i$ refere to the center of mass 
of the nucleon (the relevance of the motion of the center of mass in dipole excitations is 
one of the motivation for using non-relativistic model where the spurious contributions can 
be separated exactly) and $\Delta \alpha_E$ is a correction due to retardation effects
$\Delta \alpha_E = \langle r^2\rangle_{\rm ch}/3 M_N$ and is related to the charge mean square 
radius of the system.

In Eq.~(\ref{betam})
\begin{equation}
\mu_z = \sum_{i=1}^3\,\mu_{z,i} = \sum_{i=1}^3\,{e_i \over 2 m_i}\,\sigma_i^z
\label{muN}
\end{equation}
is the nucleon dipole magnetic moment operator, $\sigma_i$ the Pauli matrices.
$\bar \beta_M$ contains both the {\sl diamagnetic} and the {\sl paramagnetic} 
susceptibilities and, in addition, a retardation correction that can be written
\begin{equation}
\Delta \beta_M = -{3 \over 2 M}\,\langle 0|D'_z\,D'_z|0\rangle\,\,;
\label{Dbeta}
\end{equation}
Eqs.(\ref{alphae}-\ref{Dbeta}) are valid both for neutrons and protons.

From a theoretical point of view the nucleon polarizabilities received much 
attention and, in the last few years, many approaches have been developed within
complementary frameworks like constituent quark models 
\cite{constituent,drecrusso84,mtrl94}, MIT bag models \cite{mitb} and its chiral 
extensions \cite{chir}, soliton models \cite{soli1,soli2}, chiral perturbation 
theory \cite{XPT1,XPT2} and dispersion relation method \cite{disprel}.

In the present paper we want to reconsider the non-relativistic constituent quark 
model which is often believed to be inadequate to describe the static response of the
nucleon to external fields (cfr. e.g. the discussion in refs. \cite{disprel}).
As a matter of fact the constituent quark model does not incorporate 
meson degrees of freedom, and consequentely the effects of the meson cloud cannot 
be included. Therefore one cannot expect to reproduce the exprimental value of 
the electric polarizability considering quark degrees of freedom only, since the meson 
cloud should be responsable for a sizeable part of the electromagnetic reponse. However
the exact amount of the quark contribution has not been calculated by using potential
models which are fitted on the baryonic spectrum, but simply estimated within 
naive models. In the following we develop
a sum rule approach to the nucleon polarizabilities which is quite 
general to include two-body and three-body forces between the constituent quarks. 
In particular three-body forces have been recently considered in the context of 
quark models to refine the baryon mass spectrum
predictions and, in particular, to reproduce the position of the Roper resonance. 
Moreover three-body forces are strictly related to the gluon-gluon interaction 
which is one of the fundamental aspects of QCD. 
In addition we consider consistently the retardation effects arising from a 
non-relativistic approach to the electric and magnetic response.
Our results should be, therefore, considered as a precise estimate of the
quark contribution to the electric and magnetic response. 

The paper is organized in the following way: in section \ref{SRapp} we describe 
the sum rule approach to the linear response of a non relativistic system and introduce
a technique to evaluate various upper and lower bounds to the polarizability sum. The 
relation between the interquark potential model and the electric and magnetic sum rules 
is discussed in detail in section \ref{potwf} where the Isgur-Karl model, three-body
hyperradial and two-body plus three-body potential models are considered and discussed 
for both electric and magnetic excitations. The Hamiltonian is solved exactly 
by means of the Schr\"odinger equation and the wave functions of the nucleon
used to calculate the sum rules. The r\^ole of the confining, hyperfine and 
three-body forces is investigated in detail and the results summarized 
in section \ref{numres} where the comparison with experimental data is also
discussed.

\section{Sum Rule approach to nucleon polarizability}
\label{SRapp}

A quantity of the type
\begin{equation}
m_{-1}(\Theta) = \sum_{n > 0} {|\langle n|\Theta|0\rangle|^2 \over E_n - E_0}
\label{polSR}
\end{equation}
is involved in both the expressions (\ref{alphae}) and (\ref{betam}) for the 
electric polarizability and the (para) magnetic susceptibility ($\Theta$ 
corresponds to the electric dipole (\ref{DzN}) and the magnetic dipole operator 
(\ref{muN}) respectively). A direct evaluation of $m_{-1}(\Theta)$ from
Eq.~(\ref{polSR}) involves all the complications of the excitation spectrum 
of the system (energies and wave functions) and it cannot be easily estimated
truncating the excitation space \cite{notehyd}.
However one can construct upper and lower bounds 
to $m_{-1}$ \cite{mtrl94,mt96,strdf92} by using the positivity of the corresponding 
strength distribution
\begin{equation}
S_\Theta(\omega) =  \sum_{n>0} |\langle n|\Theta|0\rangle|^2\,
\delta\left(\omega - (E_n - E_0)\right)\,\,,
\label{Somega}
\end{equation}
and few ({\sl positive}) moments of such distribution ($p \geq 0$, integer)
\begin{eqnarray}
m_p(\Theta) & = & \int_0^\infty d\omega\,\omega^p\,S_\Theta(\omega) = 
\nonumber \\
& = &  \sum_{n>0} |\langle n|\Theta|0\rangle|^2\,\left(E_n - E_0\right)^p = 
\nonumber \\
& = & \langle 0|\Theta^\dagger\,(H_0 - E_0)^p\,\Theta|0\rangle
-\delta_{p0}\,|\langle 0|\Theta|0\rangle|^2\,\,.
\label{SRs}
\end{eqnarray}
The last expression has been obtained by using the closure property of the eigenstate
$|n\rangle$ and can be expressed in a simple form for the first few moments, or sum rules
( for reviews on the sum rule techniques see e.g. refs.\cite{sumrules})
\begin{eqnarray}
m_0(\Theta) & = & \langle 0|\Theta^\dagger\,\Theta|0\rangle - |\langle 0|\Theta|0\rangle|^2
\,\,,\\
m_1(\Theta) & = & {1\over 2}\langle 0|\left[\Theta^\dagger, \left[H_0,\Theta\right]\right]|0\rangle\,\,, \\
m_2(\Theta) & = & {1\over 2}\langle 0|\left\{ [\Theta^\dagger, H_0],
[H_0,\Theta]\right\}|0\rangle\,\,,\\
m_3(\Theta) & = & {1\over 2}\langle 0|\left[\left[\Theta^\dagger, H_0\right],
\left[ H_0,\left[H_0,\Theta\right]\right]\right]|0\rangle\,\,,\\
& \phantom{={1\over 2}} & {\rm etc.}\,\,.\nonumber
\end{eqnarray}
The sum rules (13 - 16) involve the ground state wave function only, and commutators 
(anticommutators) of the nucleon Hamiltonian ($H_0$) with the excitation operator 
($\Theta = D'_z\;,\mu_z$). In addition the dynamical aspects embodied in the Hamiltonian 
enter in a more and more complex way increasing the order of the sum rules and the 
explicit calculation are of increasing complexity. In the next sections we show how 
one can construct bounds to the polarizability by means of sum rules of different order.

\subsection{Lower bounds}
\label{lowerb}

In the limiting case of a strength distribution concentrated in a very narrow region
($\delta$-like distribution), the polarizability can be easily evaluated considering
sum rules of positive order only. In fact
\begin{equation}
m_{-1}(\Theta) = {m_0^2(\Theta)\over m_1(\Theta)} = \sqrt{m_1(\Theta)\,m_{-3}(\Theta)}
\;\;\; {\rm etc}\,\,.
\label{deltalike}
\end{equation}
However the strength is generally distributed over many eigenstates and its spread affects
the lower and upper moments in a different way, and the equalities (\ref{deltalike})
become rather inequalities
\begin{equation}
m_{-1}(\Theta) \geq {m_0^2(\Theta)\over m_1(\Theta)}\,\,;\;\;\; m_{-1}(\Theta) 
\leq\sqrt{m_1(\Theta)\,m_{-3}(\Theta)}\;\;\; {\rm etc}\,\,.
\label{deltaunlike}
\end{equation}
An elegant way of taking partially into account the effects of the spreading and the 
width of the strength distribution has been proposed by Dalfovo and Stringari 
\cite{strdf92} minimizing, with respect to the parameters $a$ and $b$,
the inequality
\begin{equation}
\int_0^\infty d\omega {S_\Theta(\omega)\over \omega^p}\, 
\left(1+a\,\omega+b\,\omega^2\right)^2 \geq 0\;\;\; (p\; {\rm integer})\,\,,
\label{ineq1}
\end{equation}
based on the positivity of $S_\Theta(\omega)$
\footnote{The authors of ref.\cite{strdf92} considered the case $p=1$ only.}.
One obtains
\begin{equation}
m_{-p}(\Theta) \geq {m_{-p+1}^2(\Theta)\over m_{-p+2}(\Theta)}
\,{1\over 1-{\Delta_p(\Theta)/ \Gamma_p(\Theta)}}
\label{upper1}
\end{equation}
where
\begin{equation}
\Delta_p(\Theta) = \left({m_{-p+3}\over m_{-p+2}} - 
{m_{-p+2}\over m_{-p+1}}\right)^2 \,\, ,
\end{equation}
and
\begin{equation}
\Gamma_p(\Theta) = \left[{m_{-p+4}\over m_{-p+2}} + \left({m_{-p+2}
\over m_{-p+1}}\right)^2 - 2\, {m_{-p+3}\over m_{-p+1}}\right]\,\, .
\end{equation}
A lower bound to the polarizabilty sum rule is obtained for  $p = 1$
\begin{equation}
m_{-1}(\Theta) \geq {m_0^2(\Theta)\over m_1(\Theta)}\,
{1\over 1-{\Delta_1(\Theta)/\Gamma_1(\Theta)}}\,\,.
\label{upper2}
\end{equation}

In addition to the previous relations, other bounds can be obtained by using 
Schwartz inequalities: an example involving the first odd sum rules is 
\begin{equation}
m_{-1}(\Theta) \geq {m_1^2(\Theta)\over m_3(\Theta)}
\label{upper3}
\end{equation}
based on the conditions
\begin{equation}
{m_p(\Theta)\over m_{p-2}(\Theta)} \leq {m_{p+2}(\Theta)\over m_{p}(\Theta)}
\label{ineq3}
\end{equation}
valid for the integer values of $p = 0\;,\pm1\;,\pm2$ .... etc.
Eq.~(\ref{upper3}) is obtained for $p=1$.

\subsection{Upper bounds}
\label{upperb}

Upper bounds to the polarizability are easily found considering that the 
energy transfer $\omega$ cannot be smaller than the energy difference between 
the first excited state of the system and the ground state, i.e. 
$\omega \geq E_1 - E_0$ ($=E_{10}$). The simplest way is just using the 
definition (\ref{polSR}) and applying closure
\begin{equation}
m_{-1}(\Theta) = \sum_{n > 0} {|\langle n|\Theta|0\rangle|^2 \over E_n - E_0}
\leq \sum_{n > 0} {|\langle n|\Theta|0\rangle|^2 \over E_{10}} = {m_0(\Theta) 
\over E_{10}}\,\,.
\label{lower1}
\end{equation}
Corrections to the previous bound can be considered following again ref.\cite{strdf92}:
\begin{equation}
\int_0^\infty d\omega\,{S_\Theta(\omega)\over \omega}\,\left(1+\gamma\,\omega\right)^2 \leq
\int_0^\infty d\omega\,{S_\Theta(\omega)\over E_{10}}\,\left(1+\gamma\,\omega\right)^2\,\,, 
\label{ineq5}
\end{equation}
which leads to 
\begin{eqnarray}
m_{-1}(\Theta) & \leq & {m_0(\Theta)\over E_{10}}\,
\left[1-{m_0\over m1}\,\left({m_1\over m_0}-E_{10}\right)^2\,
\left({m_2\over m_1}-E_{10}\right)^{-1}\right] = \nonumber \\
& = & {m_0(\Theta)\over E_{10}}\,\Lambda(\Theta)\,\,.
\label{lower2}
\end{eqnarray}

An even more stringent upper bound can be obtained again from Eq.(\ref{ineq1}):
in fact, for $p=2$ one obtains 
\begin{eqnarray}
m_{-1}(\Theta) & \leq & {m_1(\Theta)\,m_0(\Theta)\over m_2(\Theta)}\,
\left[1+\sqrt{\left({m_0\,m_2\over m_1^2}-1 \right)\,
\left({m_2\,m_{-2}\over m_0^2}-1\right)}\,\right] = \nonumber \\
& = & {m_1\,m_0\over m_2}\,\Sigma(\Theta)\,\,.
\label{lowerleo}
\end{eqnarray}
Note that the root argument is always positive (due to Scwartz inequalities 
and the relation (\ref{ineq1}) for $p=1$). Eq.~(\ref{lowerleo}) involves 
inverse quadratic energy weighted sum rules.

Finally, from the Schwartz inequalities (\ref{ineq3}), one can obtain 
also an upper limit for $m_{-1}$ from 
\begin{equation}
m_{-1}(\Theta) \leq \sqrt{m_{1}(\Theta)\,m_{-3}(\Theta)}\,\,,
\label{lower3}
\end{equation}
which involves the inverse cubic moment.

$m_{-2}(\Theta)$ and $m_{-3}(\Theta)$ have no closed form in terms of commutators 
and/or anticommutators; due to the quadratic and cubic energy power at the 
denominator they can be easily estimated including the first excited states 
only (in the following we will make use of the first two states). 

\section{Sum rules and interquark potential models}
\label{potwf}

The main ingredients entering the sum rules are: i) the nucleon ground state, and ii) the 
corresponding Hamiltonian 
\begin{equation}
H_0 = T + V^{(2)}_{q-q} + V^{(3)}_{q-q-q}\,\,,
\label{qhamil}
\end{equation}
which includes, in principle, two and three-body interaction terms.

\subsection{Harmonic oscillator potential model}
\label{homodel}

The most simple choice is the harmonic
oscillator two-body confining potential. It has analytic solutions and supplies a convenient 
classification scheme of the baryon resonances. If we assume a (spin-independent) harmonic 
form ($V^{(3)}_{q-q-q}=0$), 
$V^{(2)}_{q-q}=1/2\,K\,\sum_{i<j}({\bf r}_i-{\bf r}_j)^2= 
1/6\,m\,\omega^2_{\rm h.o.}\,\sum_{i<j}({\bf r}_i-{\bf r}_j)^2$, it is rather easy 
to derive the sum rules for the electric
transitions by noting that all the excited electric dipole states are degenerates at 
$E^D=E^D_{10} =1\, \hbar \omega_{\rm h.o.}$ and that
$
\left[D'_z,V^{(2)}_{q-q}+ V^{(3)}_{q-q-q}\right] = 0$.
As a consequence the linear energy-weighted sum rule assume a model independent 
form \cite{note1}
\begin{equation}
m_1(D'_z) = {1\over 2}\langle 0|\left[{D'}_z^\dagger,
\left[H_0,D'_z\right]\right]|0\rangle = {1\over 2}\langle 0|\left[{D'}_z^\dagger,
\left[T,D'_z\right]\right]|0\rangle = {e^2\over 3 m}
\label{m+1D}
\end{equation}
and the sum rules can be written
\begin{equation}
m_p(D'_z) = m_1(D'_z)\,\left(E^D\right)^{p-1} = {e^2 \over 3 m} \, \omega_{\rm h.o.}^{p-1} = 
{e^2 \over 3} \, {\alpha_{\rm h.o.}^{2p-2}\over m^p };\;\;\;\;p=0\,,\pm 1\,, \pm 2\,,
{\rm etc.}\,\,,
\label{mnho}
\end{equation}
with $\alpha_{\rm h.o.}^2=m\,\omega_{\rm h.o.}$. When $p=-1$ one derives the 
prediction of the harmonic oscillator model for the electric polarizability of neutron 
and proton \cite{drecrusso84}
\begin{equation}
\alpha_E^{\rm n,p} = 2\,m_{-1}(D'_z) = {2\over 9}\,e^2\,{M_N\over 
\alpha_{\rm h.o.}^4}\,\,.
\label{aEho}
\end{equation}
The previous equation can be expressed in a more general form as function of the 
proton charge mean square radius $\langle r_{\rm p}^2\rangle_{\rm ch} = 
1/\alpha_{\rm h.o.}^2$,
\begin{equation}
\alpha_E^{\rm n,p} = {2\over 9}\,e^2\,M_N\,
\langle r^2_{\rm p}\rangle^2_{\rm ch}\,\,,
\label{aEvar}
\end{equation}
which can be derived within a variational approach to the problem as it will be
extensively discussed in section \ref{elpol}.

The paramagnetic susceptibility vanishes in the h.o. model since the energy variation 
to flip a quark spin is zero (the interquark potential dose not depends on the spin 
degrees of freedom and therefore the $\Delta$'s mass does not differ from the nucleon mass). 
From a sum rule point of view we note that, since $[H_0,\mu_z]=0$
in the h.o. potential model, $m_1(\mu_z)=m_2(\mu_z)=m_3(\mu_z)=0$, 
while $m_0(\mu_z)={8\over9}\,\left({e\over 2 m}\right)^2$ 
(cfr. section \ref{SRMxi}) and the bounds previously introduced 
are not well defined.

\subsection{Potential models and variational approaches}
\label{potvar}

In more sophisticate versions of the interquark interaction, the confining potential 
consists of the h.o. part plus terms which removes its degeneracy. 
This is the case of the Isgur-Karl
(IK) model \cite{IKpot} which contains an unknown $U$-term added to shift the energies of 
some states. The model includes a delta-like hyperfine interaction derived from the 
one-gluon-exchange $qq$-potential. The baryonic states (obtained diagonalizing 
the Hamiltonian within a h.o. basis) result in a superposition of different $SU(6)$ 
configurations and lead to a rather good description of the baryonic spectrum. 

A sum rule approach to the nucleon polarizabilities can be developed also for that
kind of potential and we discussed results on $\alpha_E$ in a recent work \cite{mtrl94}.
However the use of sum rule technique is limited to potentials and wave functions which
are well defined and self-consistent. The IK model, making use of an unknown potential 
term, prevents explicit calculations of the sum rules. As a consequence the approach
proposed in ref.\cite{mtrl94} was limited to few sums (namely $m_1$, $m_3$ and $m_{-3}$)
and to the bounds  
\begin{equation}
2\,{m_1^2(D_z')\over m_3(D_z')}\leq
\alpha_E \leq 
2\,\sqrt{m_{1}(D_z')\,m_{-3}(D_z')}\,\,,
\label{aeIKb}
\end{equation}
because they are the only ones which still contain some informations. In fact $m_1$
is model independent and the unknown (central) $U$-potential is irrilevant. $m_{-3}$ 
can be evaluated explicitly making use of the first two excited electric 
dipole states of the nucleon spectrum ($S_{11}$ and $D_{13}$) and therefore the 
upper limit is well defined. The lower bound, even if it is written as a
combination of sum rules, defines a variational approach to $\alpha_E$ which 
assumes that the global effect of the external electric field (${\cal E}$) is 
a rigid translation of the $u$-quark wave function against the $d$-quarks without 
any additional deformation \cite{microal}. As a matter of fact variational and 
sum rule approaches are often related and we discuss this point in some detail.

\subsubsection{sum rules and wave function deformations}
\label{defosr}

Let us  assume that the wave function of the polarized nucleon can be 
obtained from the unperturbed ground state by means of a unitary transformation 
shifting charges in opposite directions in a rigid way, 
one can write a set of trial states of the form 
\begin{equation}
|\phi_{\cal E}\rangle = e^{\eta({\cal E})\,A}\,|0\rangle\,\,,
\label{trial1}
\end{equation}
$A$ is an anti-hermitian operator and $\eta({\cal E})$ the variational parameter.
Assuming $A=\left[T,D'_z\right]$ one has
\begin{equation}
|\phi_{\cal E}\rangle = e^{\eta({\cal E})\,\left[T, D'_z\right]}\,|0\rangle =
e^{-i\,\eta({\cal E})\,2\,m\,\sum_i\,e_i\,p'_{i,z}}\,|0\rangle\,\,.
\label{trial2}
\end{equation}
The states $|\phi_{\cal E}\rangle$, driven by the translation operator $p'_{z}$ 
are shifted, in the $z$-direction, in a rigid way.
For weak external fields $\eta({\cal E}) \to 0$ and one can
expand $|\phi_{\cal E}\rangle$ to evaluate the variation of
the total energy. One gets
\begin{eqnarray}
E_{\rm tot} & = & \langle \phi_{\cal E}|H_0 -D'_z {\cal
E}|\phi_{\cal E}\rangle \nonumber \\ 
& \to & \langle 0|
H_0|0\rangle -{\eta^2 \over 2} \langle 0| \left[
A, \left[H_0, A\right]\right]|0\rangle - {\cal
E} \eta \langle 0|\left[D'_z, A\right]|0\rangle
\label{Envar}
\end{eqnarray}
and the expectation value of the induced dipole defines the polarizability
\begin{equation}
\langle \phi_{\cal E}|D'_z|\phi_{\cal E}\rangle
-\langle 0|D'_z |0\rangle =  \eta  \langle 0|\left[D'_z ,A
\right]|0\rangle \equiv \alpha_E\, {\cal E}\,\, .
\label{dipolevar}
\end{equation}
Minimizing the energy  variation (\ref{Envar}) one can extract the
equilibrium value of the variational parameter
\begin{equation}
\eta = - {\cal E} {\langle 0| \left[D'_z ,
A\right]|0\rangle \over \langle 0| \left[
A,\left[H_0, A\right]\right]|0\rangle}
\label{etaeq}
\end{equation}
and substituting this expression in Eq.~(\ref{dipolevar}) one finally
obtains
\begin{eqnarray}
\alpha_E & = & - {\left(\langle 0| \left[D'_z ,
A\right]|0\rangle\right)^2 \over \langle 0|
\left[A,\left[H_0, A\right] \right]|0\rangle} = \nonumber \\
& = & {\left(\langle 0| \left[D'_z ,\left[T, D'_z \right] \right]|0
\rangle\right)^2 \over \langle 0|\left[\left[D'_z, T \right],
\left[H_0, \left[T, D'_z \right]\right] \right]|0\rangle}\,\, .
\label{varalpha}
\end{eqnarray}
Eq.~(\ref{varalpha}) represents a lower bound to the polarizability 
\cite{notevar} since the solutions of the variational
equation for $|\phi_{\cal E}\rangle$ are restricted to the wave functions obtained just
shifting the ground state charge densities of the $u$ and $d$ quarks in opposite 
directions as described by Eqs.~(\ref{trial1}) and (\ref{trial2}). However when the
Hamiltonian $H_0$ of the system in study has the property
$\left[H_0,D'_z\right]=\left[T,D'_z\right]$ because the potential commutes 
with the electric dipole operator, Eq.~(\ref{varalpha}) can be written as 
combination of sum rules
\begin{equation}
\alpha_E = {\left(\langle 0| \left[D'_z ,\left[H_0, D'_z \right] \right]|0
\rangle\right)^2 \over \langle 0|\left[\left[D'_z, H_0 \right],
\left[H_0, \left[H_0, D'_z \right]\right] \right]|0\rangle}
\equiv 2\, {m_1^2(D'_z)\over m_3(d'_z)}\,\, ,
\label{varalphas}
\end{equation}
which is valid for all the models without velocity dependent (or non-local) 
potentials, and is equivalent to the lower bound disussed in section \ref{lowerb}
and used in ref.\cite{mtrl94}.

Other possible variational approaches have a close relation with sum rules and we discuss
a second example which assumes that the wave function
of the system, under the influence of the electric field, can be approximated by means of a 
simple deformation driven by an operator $F$ such that
\begin{equation}
|\psi_{\cal E}\rangle = {1\over \sqrt {N_{\cal E}}}\,\left[1+a\,F\,\right]
\,|0\rangle\,\,.
\label{trial3}
\end{equation}
$a$ is a variational parameter, $|0\rangle$ the unperturbed ground state of the 
nucleon, and $F=\sum_i f({\bf r'}_i)$ a single particle (local) operator depending on 
the (intrinsic) quark coordinates only. $\sqrt{N_{\cal E}}$ ia a
normalization factor which ensures the condition $\langle
\psi_{\cal E}|\psi_{\cal E} \rangle = 1$ and $f({\bf r'}_i)$ a
function of ${\bf r'}_i$ to be guessed in order to introduce
physical "dipole" deformations on the wave function. We note that $F$
commutes with the dipole operator so that
$\left[F,D'_z\right]=0$, $\langle 0|F|0\rangle = 0$ because of
parity and the "dipole" character of the function
$F$, and $N_{\cal E} = 1 + a^2 \langle 0|F^2|0\rangle$.

Calculating the total energy variation one gets
\begin{eqnarray}
E_{\rm tot} & = & \langle \psi_{\cal E}|H_0 - D'_z {\cal
E}|\psi_{\cal E}\rangle \nonumber \\ 
& \to & \left(1 - a^2 \langle 0|F^2|0\rangle \right)
\left(\langle 0|H_0|0\rangle +a^2 \langle 0|F\,H_0\,F|0\rangle 
-2 a {\cal E} \langle 0|F\,D'_z|0\rangle \right)\nonumber \\
& = & E_0 + {a^2 \over 2} \langle 0| \left[F, \left[H_0,
F\right]\right]|0\rangle - 2 a {\cal E} \langle 0|F\,D'_z|0\rangle \,\, ,
\label{Entrial3}
\end{eqnarray}
and
\begin{equation}
\langle \phi_{\cal E}|D'_z|\phi_{\cal E}\rangle -
\langle 0|D'_z|0\rangle = 2 a \langle 0|F\,D'_z|0\rangle =
\alpha_E\,{\cal E}\,\, .
\label{dipoletrial3}
\end{equation}
Minimizing (\ref{Entrial3}) and substituting in (\ref{dipoletrial3}) the 
obtained equilibrium value of $a$, one has
\begin{equation}
\alpha_E = 4\, {\left(\langle 0|F\,D'_z|0\rangle\right)^2 \over 
\langle 0|\left[F,\left[H_0, F\right] \right]|0\rangle}\,\, .
\label{alphaEtrial3}
\end{equation}
The previous expression is similar to Eqs.~(\ref{varalpha}), (\ref{varalphas}) and can 
be evaluated by using canonical commutation relations\footnote{
The two example given in this section could be unified within the unitary
transformation formalism of Eq.~(\ref{trial1}), considering the operator $A$ as a many-body
operator satisfying the relations $\langle 0|\left[A,D'_z\right]|0\rangle = \langle 0|F\,D'_z|0\rangle$ and $\langle 0|\left[A,\left[H_0, A\right]\right]|0\rangle
= \langle 0|\left[F,\left[H_0, F\right] \right]|0\rangle$.}. 

A quite natural and simple choice for $F$ is $F = D'_z$. In this case 
\begin{equation}
\alpha_E = 4\, {\left(\langle 0|D'_z\,D'_z|0\rangle\right)^2 \over 
\langle 0|\left[D'_z,\left[H_0, D'_z\right] \right]|0\rangle} =
2\, {m_0^2(D'_z) \over m_1(D'_z) }\,\,,
\label{trial3SR}
\end{equation}
and a simplified lower bound is found. It is rather similar to the bound 
(\ref{upper2}) and can be obtained from the constrain (\ref{ineq1}) in the
limit $p=1$ and $b=0$.

\subsection{Hyperradial potentials and three-body forces}

Just at the opposite side of the simple two-body harmonic potential, one can 
locate the three-body force model (TBM) recently proposed by Ferraris, Giannini, 
Pizzo, Santopinto and Tiator \cite{TBM95}. The model assumes that the interquark 
potential can be written as an hypercentral potential, i.e. a potential depending 
on the hyperradius only. The idea of hyperradius ($\xi$) is introduced in 
the so called hyperspherical formalism \cite{hyper} together with the hyperangle 
($\phi$), to define the hyperspherical set of coordinates in a six-dimensional 
space: $\Omega_\rho$, $\Omega_\lambda$, $\xi$ and $\phi$, with
$\xi = \sqrt{\rho^2 + \lambda^2}$, $\phi = \arctan\left({\rho \over \lambda}\right)$
and $\rho$ and $\lambda$ are the absolute values of the Jacobi coordinates
$\vec \rho = ({\bf r}_1-{\bf r}_2)/\sqrt{2}$, and 
$\vec \lambda = ({\bf r}_1+{\bf r}_2 -2\,{\bf r}_3)/\sqrt{6}$.

Under the assumption of an hypercentral potential the intrinsic Hamiltonian $H_0$ 
can be written 
\begin{equation}
H_0 = - {1\over 2\,m}\left(\nabla^2_\rho+\nabla^2_\lambda\right)+V^{(3)}(\xi)=
- {1\over 2\,m}\left({d^2\over d\xi^2}+{5\over \xi}\,{d\over d\xi}+
{L^2(\Omega)\over \xi^2}\right)
+V^{(3)}(\xi)\,\,,
\label{hyperHo}
\end{equation}
where $L^2(\Omega)/\xi^2$ is the analogous, in six dimensions, of the 
three-dimensional centrifugal barrier and $\Omega$ embodies 
$\Omega_\rho, \Omega_\lambda, \phi$. The eigenfunctions of the grand-angular 
operator $L^2$ are called hyperspherical harmonics and denoted by $Y_{[\gamma]}(\Omega)$.
They are written as products of spherical harmonics in $\Omega_\rho$ and $\Omega_\lambda$
with angular momentum $l_\rho$ and $l_\lambda$ (corresponding to the coordinates $\vec \rho$
and $\vec \lambda$) and Jacobi polynomials in the hyperangle $\phi$ \cite{hyper}. They 
form a complete orthogonal basis in the space of the five-dimensional functions of 
$\Omega_\rho$, $\Omega_\lambda$ and $\phi$. The eigenvalues
of $L^2$ are $-\gamma\,(\gamma+4)$, $\gamma$ are called the grand-angular quantum 
numbers and are given by $\gamma = 2\,k+l_\rho+l_\lambda$ with $k$ integer and non-negative.
As long as the potential is hypercentral the complete wave function $\Psi(\xi,\Omega)$ 
can separated as $\Psi(\xi,\Omega)=\psi_{\nu,\gamma}(\xi)\,Y_{[\gamma]}(\Omega)$ and 
the hyperradial wave function $\psi_{\nu,\gamma}(\xi)$ is solution of a
Scr\"odinger equation. For fixed value of $\gamma$, the label $\nu$ indicates the number 
($\nu+1$) of nodes in the wave function. 

The hypercentral character of the potential means (in general) that the interquark 
interaction has a genuine three-body character, in the sense that the coordinates 
of a specific pair cannot be disentangled from the third one \cite{note3}. 
The idea of a three-quark force is related to the non-abelian character of QCD, in 
particular to the existence of a direct gluon-gluon interaction which represents a 
fundamental justification for the introduction of three-body forces \cite{tbQCD}. From 
a more phenomenological point of view it has been argued \cite{roper1,roper2} that the 
contribution of a three-body force is of some help in solving the "Roper puzzle" 
\cite{note4} giving a consistent explanation to its relative position in the spectrum. 

\vspace{2mm}

For a given hypercentral potential the eigenvalue equation has to be solved numerically 
obtaining energies and hypercentral wave functions which, depending on the hyperradius 
only, result to be completely symmetric. The correct global symmetry of the state is 
obtained combining $\psi_{\nu,\gamma}(\xi)$ with the appropriate  
hyperangular, angular, spin and isospin wave functions \cite{ESthesis}. 
In particular
the nucleon wave function is written as 
\begin{equation}
|N,J^P={1/2}^+\rangle = \psi_{00}(\xi)\,Y_{[0,S]}^{(0,0)}(\Omega)\,\chi_S(1/2;1/2)\,\,,
\label{hyperNuc}
\end{equation}
where $\chi_S(1/2;1/2)$ is the symmetric $SU(6)$ combination of spin and isospin wave 
functions of the three quarks, and we have explicitly indicated the symmetry required for 
the hypershperical wave function and the total angular momentum 
$\vec L= \vec l_\rho + \vec l_\lambda$: $Y_{[\gamma,{\rm symmetry}]}^{(L,M)}$.
For any given $SU(6)$-configuration only $\psi_{\nu,\gamma}(\xi)$ will depend on 
the potential and we can write general results for all the sum rules when 
$H_0=T+V^{(3)}(\xi)$ \cite{EBthesis}.

\subsubsection{sum rules for the electric dipole excitations}
\label{SRExi}

\vspace{2mm}

\noindent i) $m_0(D'_z)$:

\begin{eqnarray}
m_0(D'_z) & = & \langle\sum_i\,e_i^2\,{z'_i}^2\rangle + 
\langle\sum_{i\neq j}\,e_i\,e_j\,\,z'_i\,z'_j\rangle = \nonumber\\
& = & {1\over 9}\,e^2\,\int\,d\xi\,\xi^7\,|\psi_{00}(\xi)|^2 = 
{1\over 3}\,e^2\,\langle r_{\rm p}^2\rangle_{\rm ch}
\label{m0Dhyp}
\end{eqnarray}
both for protons and neutrons. The non-energy-weighted sum rule is therefore 
proportional to the mean square mass radius of the nucleon, which, at least as long 
as the nucleon wave function belongs to a $SU(6)$ multiplet, is also the proton 
charge radius.

\vspace{2mm}

\noindent ii) $m_1(D') = {e^2/ 3 m}$
remains unmodified for all the velocity independent potentials as already discussed (cfr. Eq.~(\ref{m+1D})) and does not distinguish neutron and proton. Its model independence
is based on the property $[V^{(3)}(\xi), D'_z]=0$

\vspace{2mm}

\noindent iii) $m_2(D'_z)$

\begin{eqnarray}
m_2(D'_z) & = & -{1\over 2 m^2}\,\langle\,{1\over 3}(e_1+e_2-2e_3)^2\,\nabla^z_\lambda\,
\nabla^z_\lambda+(e_1-e_2)^2\,\nabla^z_\rho\,\nabla^z_\rho + \nonumber \\
& + & {2\over\sqrt 3}\,(e_1+e_2-2e_3)\,(e_1-e_2)\,\nabla^z_\rho\,\nabla^z_\lambda\rangle =
\nonumber \\
& = & {2\,e^2\over 9 m}\,\langle T\, \rangle = -{e^2\over 9 m^2}\,\int d\xi\,\xi^5\,\,
\psi^*_{00}(\xi)\,\left( {d^2 \over d\xi^2}+
{5\over \xi}\,{d \over d\xi}\right)\,\psi_{00}(\xi)\,\,.
\label{m2Dhyp}
\end{eqnarray}
$m_2$ does not depend on the interquark interaction explicitly, however is proportional 
to the mean kinetic energy of the system and measures the presence of high momentum 
components in the nucleon wave functions. If the potential is highly confining the 
high momentum components are larger and the sum increases.

\vspace{2mm}

\noindent iv) $m_3(D'_z)$

\begin{eqnarray}
m_3(D'_z) & = & {1\over 2 m^2}\,\langle{1\over 6}\,(e_1+e_2-2e_3)^2\,(\nabla^z_\lambda\,
\nabla^z_\lambda\,V^{(3)}(\xi)) +  \nonumber \\
& + & {1\over 2}\,(e_1-e_2)^2\, (\nabla^z_\rho\,
\nabla^z_\rho\,V^{(3)}(\xi))\rangle + \nonumber\\
& + & {1\over \sqrt{3}}\,(e_1+e_2-2e_3)\,(e_1-e_2)\,(\nabla^z_\lambda\,
\nabla^z_\rho\,V^{(3)}(\xi))\rangle = \nonumber\\
& = & {e^2\over 18 m^2}\,\int d\xi\,\xi^5\,
|\psi_{00}(\xi)|^2\, \left({d^2 V^{(3)}(\xi)\over d\xi^2}+
{5\over \xi}\,{d V^{(3)}(\xi)\over d\xi}\right)\,\,.
\label{m3Dhyp}
\end{eqnarray}
The three-times energy-weighted sum is particularly interesting because it 
depends crucially on the potential model and on its derivatives. 

\subsubsection{sum rules for magnetic dipole excitations}
\label{SRMxi}

Magnetic excitations induced by the dipole operator (\ref{muN}) are easily understood 
in the case of spin-independent interactions. In fact the magnetic operator commutes 
with the nucleon Hamiltonian $[T+V^{(3)}\xi), \mu_z]=0$ and the total strenght is 
concetrated at a value of the excitation energy equal to the ground state and one gets:

\vspace{2mm}

\noindent i) $m_0(\mu_z)$

\begin{eqnarray}
m_0(\mu_z) & = & {1\over (2\,m)^2}\,\left[\langle \sum_{i\neq j}e_i\,e_j\,\sigma^z_i\,
\sigma^z_j \rangle +\langle \sum_i e_i^2\rangle - |\langle \sum_i e_i\,\sigma_i^z
\rangle|^2 \right]\nonumber \\
& = & \left({e\over 2\,m}\right)^2\,\left[{8\over 9} + 1 -1\right]
 = {8\over 9}\,\mu_0^2\,\,.
\label{m0Mhyp}
\end{eqnarray}
The result (\ref{m0Mhyp}) is valid for both protons and neutrons (in the neutron case 
the indidual contributions in parenthesis become $2/3 +2/3-4/9$), and we have defined
$\mu_0=e/2\,m$.

\vspace{2mm}

\noindent ii) {\it higher sum rules}:

\begin{equation}
m_1(\mu_z)=m_2(\mu_z)=m_3(\mu_z)=0
\label{mpMhyp}
\end{equation}
vanish because we did not consider the effects due to hyperfine interaction. 
In the following we discuss a variety of hypercentral potential investigated in 
ref.\cite{TBM95}, and to this hand we discuss first the role of the spin-spin 
interaction terms.

\subsection{The role of the hyperfine interaction}
\label{hyperfine}

The authors of the ref.\cite{TBM95} include a perturbative Hyperfine interaction in the 
three-body Hamiltonian in order to reproduce the correct $N$-$\Delta$ mass splitting.
Such contribution is seen as the non-relativistic reduction of the one-gluon-exchange
interaction \cite{RujGeoGla75}. We restrict the Hyperfine contribution to the dominant 
spin-spin zero-range term
\begin{equation}
V^{(2)}_{\rm Hyp}({\bf r}_{12}) = {2\,\alpha_S \over 3\,m^2}\,{8\,\pi \over 3}\,{\bf S}_1
\cdot{\bf S}_2\,\delta({\bf r}_{12})
\label{HyperfSS}
\end{equation}
neglecting the (small) tensor contribution. The effective coupling constant $\alpha_S$
is fixed by the $N$-$\Delta$ mass splitting
\begin{equation}
M_{\Delta}-M_{N} = {\sqrt{2}\over 3}{\alpha_S \over \,m^2}\,{8 \over \pi}\,
\int d\xi\,\xi^2\,|\psi_{00}(\xi)|^2\,\,.
\label{MD-N}
\end{equation}

The additional contributions due to the Hyperfine potential (\ref{HyperfSS}) prevents 
the magnetic dipole sums to vanish, and modifies the energy-weighted electric dipole sums.

\subsubsection{sum rules for the elctric dipole excitations}
\label{HypDSR}

\noindent i) $m_1(D'_z)$ and $m_2(D'_z)$ remain unmodified since 
$\left[D'_z, V^{(2)}_{\rm Hyp}\right] = 0 $

\vspace{2mm}

\noindent ii) $m_3(D'_z)$ gains an additional contribution:

\begin{eqnarray}
\left. m_3(D'_z)\right|_{\rm Hyp} & = & {\sqrt{2} \pi \over 9}\,{\alpha_S \over m^4}\,
\langle(e_1-e_2)^2\,{\bf S}_1 \cdot {\bf S}_2\,[\nabla_\rho^2\,\delta({\vec \rho})]\rangle
\nonumber \\
& = & - {1\over 3}\,{e^2\over m^2}\,\left(M_{\Delta}-M_N\right)
\,{\int d\xi\,\xi\,\psi_{00}(\xi)\,\psi'_{00}(\xi) \over 
\int d\xi\,\xi^2\,|\psi_{00}(\xi)|^2}\,\,,
\label{m3xHyp}
\end{eqnarray}
which has to be added incoherently to the Eq. (\ref{m3Dhyp}). 
In Eq. (\ref{m3xHyp}) $\psi'_{00}(\xi)$ indicates the total derivative 
of the wave function and the relation  (\ref{MD-N}) has been used \cite{notedelta}. 

\subsubsection{sum rules for magnetic dipole excitations}
\label{HypMSR}

The crucial role of the Hyperfine interaction for the magnetic energy-weighted sum rules 
has been already pointed out in the previous sections by stressing the fact that these
sum rules simply vanish if the spin-spin term (\ref{HyperfSS}) is neglected. 
Hyperfine potential is, 
however, a two-body operator and the commutators with the magnetic dipole moment (\ref{muN}) 
are not as simple as the electric case. For simplicity we will restrict to the linear 
energy-weighted sum rule which, differently from $m_1(D'_z)$, is not model independent and 
embodies the interesting aspects of the interaction-dependence we recognize in $m_3(D'_z)$.

As long as the eigenstates of the hypercentral potential are assumed to belong to
a specific $SU(6)$ multiplet, the magnetic operator can excite one state only: the 
$\Delta_{33}$ resonance. Within such simple $N$-$\Delta$ excitation model the 
magnetic susceptibility is given by
\begin{equation}
\beta_M = {|\langle \Delta|\mu_z|N\rangle|^2 \over M_\Delta - M_N} = 
{16\over 9}\,\mu_0^2\,{1\over M_\Delta - M_N} \approx 8.7\cdot 10^{-4}\,
{\rm fm}^3\,\,.
\label{bNDmodel}
\end{equation}
The sum rule approach cannot add any information to this simple scheme
as it can be demonstrated calculating the first moments of the magnetic 
strength distribution:

\vspace{2mm}

\noindent i) $m_0(\mu_z)$ has the form (\ref{m0Mhyp}). 

\vspace{2mm}

\noindent ii) $m_1(\mu_z)$

\begin{eqnarray}
m_1(\mu_z) & = & - {12 \over m^2}\,\langle (e_1-e_2)^2\,\left({\bf S}_1 \cdot 
{\bf S}_2 - S_1^z\, S_2^z\right)\,{2\pi\over 9}\,{\alpha_S\over m^2}\,
\delta({\bf r}_{12}) \rangle \nonumber \\
& = & \alpha_S\,{1\over 27 \sqrt{2}}{e^2 \over \,m^4}\,{16 \over \pi}\,
\int d\xi\,\xi^2\,|\psi_{00}(\xi)|^2 \nonumber \\
& = & {8\over 9}\,\mu_0^2\,\left(M_{\Delta}-M_{N}\right)\,\,,
\label{m1MHyp}
\end{eqnarray}
where Eq.~(\ref{MD-N}) has been used.

By using the bounds involving $m_1$, $m_0$ and $E_{10}^M = M_\Delta - M_N$
one obtains
\begin{equation}
2\,{m_0^2(\mu_z)\over m_1(\mu_z)} = {16\over 9}\,\mu_0^2\,{1\over M_\Delta - M_N} 
\leq \beta_M \leq 
2\,{m_0(\mu_z)\over E^M_{10}} = {16\over 9}\,\mu_0^2\,{1\over M_\Delta - M_N} 
\,\,,
\label{ulb-1}
\end{equation}
and the two limits coincides with the result (\ref{bNDmodel}) as expected.

An estimate of $\beta_M$ different from the previous result comes from $SU(6)$ 
breaking components in the nucleon wave function. In the next section we discuss 
a potential model whose nucleon wave function results in a superposition of 
different $SU(6)$ configurations.

\subsection{Two-body + three-body interactions}
\label{twothreeV}

The pathologic aspect of non-relativistic quark models involving confining potentials 
plus (at least part of ) the one-gluon-exchange interaction remains the impossibility to
predict the masses of the first (negative parity) excitations in the baryonic
octet and decuplet correctly. The authors of ref.\cite{roper2} proposed to add a 
phenomenological three-quark force getting a correct spectrum up to excitation energy 
of $0.7$ GeV. The 
analysis of the mesonic $q\bar q$ and baryonic $qqq$ spectrum leads to a two-body 
potential containing confining, Coulomb-like and spin-spin $qq$ interaction of the 
form \cite{BCNinter}
\begin{equation}
V^{(2)}_{q-q} = {1\over 2}\left(- {\kappa \over r_{12}} +
{r_{12}\over a^2} + 4\,{\kappa_\sigma \over m^2}\,{\exp(-r_{12}/r_0) \over r_0^2\,r_{12}}\,
{\bf S}_1 \cdot {\bf S}_2 - D\right)\,\,,
\label{VqqBC}
\end{equation}
where the Yukawa form of the spin-spin term replaces the delta contact interaction of the 
OGE potential (\ref{HyperfSS}) to avoid an unbounded spectrum when the Scr\"odinger 
equation is solved.

The three-body term suggested by Cano {\it et al.} in ref.\cite{roper2}
\begin{equation}
V^{(3)}_{q-q-q} = V_0 \, \exp \left( - \sum_{i<j}\,{r^2_{ij} \over \lambda_0^2}\right)
= V_0 \, \exp \left( -3 \xi^2/\lambda_0^2 \right)
\label{Vqqqca}
\end{equation}
is identified as $V^{(3)}_{III}$ by the same authors and assumes a simple form
within the hyperspherical formalism.
The parameters of the interaction can be found in table 1 of their paper and 
in the following we discuss the results of such quite recent version of
the QCD inspired potential including two- and thre-body contributions.

The solutions of the Schr\"odinger equation can be expanded on the 
hyperspherical basis and this is the actual procedure followed in ref.\cite{roper2}.
In practical calculations only two terms have been retained and the nucleon wave function
can be written:
\begin{eqnarray}
& & |N,J^P ={1/2}^+\rangle = \Psi_1(\xi)\,
Y^{(0,0)}_{[0,{\rm S}]}\,\chi_{[{\rm S}]}(1/2;1/2) + \nonumber \\
& + & \Psi_3(\xi)\,{1\over \sqrt 2}\,\left[
Y^{(0,0)}_{[2,{\rm MS}]}\,\chi_{[{\rm MS}]}(1/2;1/2) + 
Y^{(0,0)}_{[2,{\rm MA}]}\,\chi_{[{\rm MA}]}(1/2;1/2)\right]\,,
\label{wfcano}
\end{eqnarray}
where the second term of the expansion is clearly an $SU(6)$-breaking contribution,
and $\chi_{[{\rm MS}]}(1/2;1/2)$, ($\chi_{[{\rm MA}]}(1/2;1/2)$)
are the mixed-symmetric (antisymmetric) $SU(6)$ combinations of spin and isospin
wave functions of the three quarks.
Owing to the $SU(6)$-breaking term in the nucleon wave function, some results
of the pure hyperradial potentials are modified. For istance the charge root mean
square radius of the neutron is not vanishing and the values depend on the contributions
coming from the three-body part of the potential. 

\subsubsection{sum rules for electric dipole excitations}
\label{canoESR}

\vspace{2mm}

\noindent i) $m_0(D'_z)$:

\begin{eqnarray}
m_0(D'_z)
& = & {1\over 9}\,e^2\,\left\{\int\,d\xi\,\xi^7\,\left[|\Psi_1(\xi)|^2+|\Psi_3(\xi)|^2\right] 
- {1\over \sqrt 2} \int\,d\xi\,\xi^7\,\Psi_1(\xi)\,\Psi_3(\xi)\right\}\nonumber\\
& = & {1\over 3}\,e^2\,\langle r^2_{\rm p}\rangle_{\rm ch} +
{2\over 3}\,e^2\,\langle r^2_{\rm n}\rangle_{\rm ch}\,\,,
\label{m0Dcano}
\end{eqnarray}
both for protons and neutrons. The non-energy-weighted sum rule, 
therefore, is no longer proportional to the charge mean square radius of 
the proton when $SU(6)$ breaking component in the nucleon wave 
function are taken into account.

\vspace{2mm}

\noindent ii) $m_1(D'_z) = {e^2/ 3 m}$
again remains unmodified because 
$[V ^{(2)}_{q-q}(\xi)+V^{(3)}_{q-q-q}(\xi), D'_z]=0$
as already discussed and does not distinguish neutron and proton. 

\vspace{2mm}

\noindent iii) $m_2(D'_z)$ is no longer proportional to the mean 
kinetic energy and contains an additional term:
\begin{equation}
m_2(D'_z) = {2\,e^2\over 9 m}\,\langle T\, \rangle + {e^2\over 9 m^2}\,
{1\over \sqrt{2}} \int d\xi\,\xi^5\,\,
\Psi^*_{3}(\xi)\,\left( {d^2 \over d\xi^2} -
{1\over \xi}\,{d \over d\xi}\right)\,\Psi_{1}(\xi)\,\,;
\label{m2Dcano}
\end{equation}

\vspace{2mm}

\noindent iv) $m_3(D'_z)$

\noindent Both $V^{(2)}_{q-q}$ and $V^{(3)}_{q-q-q}$ contribute to the sum rule 
which can be written 
$m_3(D'_z)=\left. m_3(D'_z)\right|_{q-q}+\left. m_3(D'_z)\right|_{q-q-q}$, where

\begin{equation}
\left. m_3(D'_z)\right|_{q-q} = {1\over 4 m^2}\,\langle(e_1-e_2)^2\,\vec \nabla^2_{r_{12}}
\left[- {\kappa_C \over r_{12}} +
{r_{12}\over a^2} + 4\,{\kappa_\sigma \over m^2}\,{\exp(-r_{12}/r_0) \over r_0^2\,r_{12}}\,
{\bf S}_1 \cdot {\bf S}_2\right] \rangle \,\,,
\label{m3Dcanoqq}
\end{equation}

and the three-body $\left.m_3(D'_z)\right|_{q-q-q}$ is again given by Eq.~(\ref{m3xHyp})
with the obvious replacement $V^{(3)}(\xi)) \to V^{(3)}_{q-q-q}(\xi)) $.

\subsubsection{sum rules for magnetic excitations}

The role of $SU(6)$ breaking components in the nucleon wave function (\ref{wfcano})
is particularly relevant for the magnetic excitations because the operator $\mu_z$ can 
now  mix different $SU(6)$ configurations. We discuss the simple bounds which involve
$m_0$, $m_1$ and $E_{10}^M$ only. One gets

\vspace{2mm}

\noindent i) $m_0(\mu_z)$

\begin{equation}
m_0(\mu_z) = \mu_0^2\,\left[1 + {8\over 9} P_1 - \left( P_1 + {1\over 3} P_3\right)^2\right]
\,\,,
\label{m0Mqqq}
\end{equation}
where $P_\alpha = \int d\xi\,\xi^5\,|\psi_\alpha(\xi)|^2$ and one recovers the result
(\ref{m0Mhyp}) in the limit $P_1=1$ and $P_3=0$.
In the neutron case the sum rule reads
\begin{eqnarray}
m_0(\mu_z) = \mu_0^2\,{2\over 3}\left[1 + \left( P_1 + {1\over 3} P_3\right) -
{2\over 3} P_1^2 \right]
\,\,.
\label{m0Mqqqn}
\end{eqnarray}
However, despite the different form of the sum rules the numerical values are
quite close each other and do not produce differences in the estimates of the
$\beta_M$ for protons and neutrons.

\vspace{2mm}

\noindent ii) $m_1(\mu_z)$

\begin{equation}
m_1(\mu_z) = - {12 \over m^2}\,\langle (e_1-e_2)^2\,\left({\bf S}_1 \cdot 
{\bf S}_2 - S_1^z\, S_2^z\right)\,{1\over 4}\,
{\kappa_\sigma \over m^2}\,{\exp(-r_{12}/r_0) \over r_0^2\,r_{12}}\rangle \,\,.
\label{m1Mqqq}
\end{equation}

\section{Numerical results and discussion}
\label{numres}

We discuss, first, numerical results for a variety of hyperradial potentials 
introduced by Ferraris {\it et al.} \cite{TBM95}, namely

\begin{eqnarray}
V^{(3)}_1(\xi) & = & - {\tau\over \xi} + k_l\,\xi \label{V1}\\
V^{(3)}_2(\xi) & = & - {\tau\over \xi} + k_l\,\xi + {b \over \xi^2} \label{V2}\\
V^{(3)}_3(\xi) & = & - {\tau\over \xi} + k_l\,\xi + {b \over \xi^2} + c\,{\rm log} \xi\label{V3}\\
V^{(3)}_4(\xi) & = & - {\tau\over \xi} \label{V4}\\
V^{(3)}_5(\xi) & = & - {\tau\over \xi} + {b\over \xi^2} \label{V5}
\end{eqnarray}
whose parameters are summarized, for convenience, in table \ref{tableVx}. The potentials 
$V_1$ - $V_3$ have been fitted on the baryon mass spectrum, the hypercoulomb 
parametrization (\ref{V4}) has been considered as check of numerical calculations, while the 
version (\ref{V5}) has been introduced in ref.\cite{TBM95} because it reproduces the 
electric dipole form factor and root mean square radius of the proton. In the same 
table the predictions of the mass radius are also presented. 

\begin{table}

\caption{Parameters of the potentials (\ref{V1}) - (\ref{V5}). 
In the last coulomn the predicted mass radii.}

\begin{tabular}{ccccccccccccccccccccccccccccccccccccccccccccccccccccccccccccccccccccccccccccc}

& ${\rm potentials}$ & $\tau$ & $k_l$ & $b$ & $c$ &
$\langle r^2\rangle$ & \\
& & [u] & [fm$^{-2}$] & [fm] & [fm$^{-1}$] & [fm$^{2}$] & \\
\hline 
&$V^{(3)}_1$ & $4.59$ & $1.61$ & $-$ & $-$ & $(0.516)^2$ & \\
&$V^{(3)}_2$ & $2.50$ & $1.14$ & $-0.80$ & $-$ & $(0.483)^2$ & \\
&$V^{(3)}_3$ & $2.50$ & $1.14$ & $-0.80$ & $0.1$ & $(0.475)^2$ & \\
&$V^{(3)}_4$ & $6.39$ & $-$ & $-$ & $-$ & $(0.462)^2$ & \\
&$V^{(3)}_5$ & $1.78$ & $-$ & $-0.78$ & $-$ & $(0.88)^2$ & \\
\label{tableVx}
\end{tabular}
\end{table}

\subsection{Electric dipole polarizability}
\label{elpol}

\subsubsection{results of three-body force models (TBM)}

The electric dipole sum rules are shown in table \ref{tableSRx} for the hyperradial 
interactions (\ref{V1}) - (\ref{V5}), and the corresponding lower and upper limits in tables
\ref{tableLBx}, and \ref{tableUBx}. The smaller is the predicted radius of the system, 
the smaller is the non-energy-weighted sum and the larger the $m_2$ moment. In fact $m_0(D'_z)$
is proportional to the proton charge radius and $m_2(D'_z)$ to the mean kinetic energy which
is larger in smaller systems because of the indetermination principle. The role of the 
hyperfine interaction is clearly seen in $m_3(D'_z)$ (cfr. table \ref{tableSRx}).
A large part of the sum comes from the spin-spin interaction and the lower limit
$2\,m_1^2/m_3$ can become significantly small (cfr. table \ref{tableLBx}) loosing a 
direct connection with reasonable values of the polarizability. One can 
conclude that the charge deformation induced by the external electric field on the quark 
distribution is far from beeing approximated by a rigid oscillation of $u$ and $d$
charge densities (opposite in phase) as assumed in the relations (\ref{varalpha}), 
and (\ref{varalphas}). The inclusion of hyperfine contributions to the $m_3$ sum 
lowers the polarizability by few percent enlarging the difference between the upper 
and lower limits.

\begin{table}

\caption{Electric dipole sum rule values for the potential models (\ref{V1})-(\ref{V5})
(cfr. also table \ref{tableVx}). For the $m_3$ sum the contributions coming from the 
hyperradial part of the potential is shown, the total result contains, in addition, the
Hyperfine contribution coming from the interaction term (\ref{HyperfSS}).}

\begin{tabular}{ccccccccccccccccccccccccccccccccccccccccccccccccccccccccccccccccccccccccccccc}

& ${\rm potentials}$ & $m_0(D'_z)$ & $m_1(D'_z)$ & $m_2(D'_z)$ & $m_3(D'_z)|_{V^{(3)}}$ &
$m_3(D'_z)|_{\rm tot}$ & \\
& & [fm$^2$] & [fm] & [u] & [fm$^{-1}$] & [fm$^{-1}$] & \\
\hline 
&$V^{(3)}_1$ & $0.0006$ & $0.0015$ & $0.0040$ & $0.0137$ & $0.0278$ & \\
&$V^{(3)}_2$ & $0.0006$ & $0.0015$ & $0.0056$ & $0.0760$ & $0.2085$ & \\
&$V^{(3)}_3$ & $0.0005$ & $0.0015$ & $0.0058$ & $0.0799$ & $0.2149$ & \\
&$V^{(3)}_4$ & $0.0005$ & $0.0015$ & $0.0053$ & $0.0274$ & $0.0498$ & \\
&$V^{(3)}_5$ & $0.0019$ & $0.0015$ & $0.0018$ & $0.0098$ & $0.0602$ & \\
\label{tableSRx}
\end{tabular}
\end{table}

\begin{table}

\caption{Lower bounds to the electric polarizability as predicted by the potential models
(\ref{V1})-(\ref{V5}). In parenthesis the numerical results obtained neglecting the 
Hyperfine contribution in $m_3$ (cfr. table \ref{tableSRx}). The polarizability is 
expressed in $10^{-4}$ fm$^3$.}

\begin{tabular}{ccccccccccccccccccccccccccccccccccccccccccccccccccccccccccccccccccccccccccccc}

& ${\rm potentials}$ & $2\,{m_1^2(D'_z)\over m_3(D'_z)}$ &
$2\,{m_0^2(D'_z)\over m_1(D'_z)}$ & 
$2\,{m_0^2(D'_z)\over m_1(D'_z)(1-\Delta/\Gamma)}$ & \\
\hline 
& $V^{(3)}_1$ & $1.69$ ($3.44$) & $5.48$ & $5.50$ ($5.60$) & \\
& $V^{(3)}_2$ & $0.23$ ($0.62$) & $4.19$ & $4.22$ ($4.29$) & \\
& $V^{(3)}_3$ & $0.22$ ($0.59$) & $3.94$ & $3.97$ ($4.04$) & \\
& $V^{(3)}_4$ & $0.94$ ($1.72$) & $3.51$ & $3.55$ ($3.65$) & \\
& $V^{(3)}_5$ & $0.78$ ($4.79$) & $45.7$ & $45.9$ ($47.1$) & \\
\label{tableLBx}
\end{tabular}
\end{table}

Looking at the comparison between the simple lower bound 
\begin{equation}
\alpha_E^{\rm p,n} = 2\,{m_0^2(D'_z)\over m_1(D'_z)} = 
{2\over 9}\,e^2\,\left(3\,m\right)\,\langle r^2_{\rm p}\rangle_{\rm ch}^2
\label{sbound}
\end{equation}
shown in table \ref{tableLBx} and the results of table \ref{tablealphaEx} one can see 
that the approximation (\ref{sbound}) is rather good for all the potential models
suggesting that Eq.~(\ref{trial3}) with $F\equiv D'_z$ represents a  more reliable 
nucleon charge deformation induced by the electric field.

\begin{table}

\caption{Upper bounds to the electric polarizability as predicted by the potential models
(\ref{V1})-(\ref{V5}). The polarizability is expressed in $10^{-4}$ fm$^3$.}

\begin{tabular}{ccccccccccccccccccccccccccccccccccccccccccccccccccccccccccccccccccccccccccccc}

& potentials & ${m_1(D'z)\,m_0(D'z)\over m_2(D'z)}\,\Sigma$ &
$2\,{m_0(D'_z)\over E^D_{10}}$ &
$2\,{m_0(D'_z)\over E^D_{10}}\,\Lambda$ & $2\,\sqrt{m_1(D'_z)\,m_{-3}(D'_z)}$ & \\
\hline
& $V^{(3)}_1$ & $5.65$ & $5.86$ & $5.71$ & $5.88$ & \\
& $V^{(3)}_2$ & $4.53$ & $5.08$ & $4.78$ & $5.04$ & \\
& $V^{(3)}_3$ & $4.27$ & $4.79$ & $4.51$ & $4.75$ & \\
& $V^{(3)}_4$ & $3.64$ & $4.09$ & $3.82$ & $3.95$ & \\
& $V^{(3)}_5$ & $46.5$ & $66.2$ & $58.1$ & $56.0$ & \\
\label{tableUBx}
\end{tabular}
\end{table}

\begin{table}

\caption{Bounds to the electic polarizability as predicted by the potential models
(\ref{V1})-(\ref{V5}). The results obtained neglecting Hyperfine contributions 
are also shown. The polarizability is expressed in $10^{-4}$ fm$^3$.}

\begin{tabular}{ccccccccccccccccccccccccccccccccccccccccccccccccccccccccccccccccccccccccccccc}

& potentials & $V^{(3)}(\xi)$ [only] & $V^{(3)}(\xi)+V^{(2)}_{\rm Hyp}$ & \\
\hline
& $V^{(3)}_1$ & $\alpha_E^{\rm p,n}=5.65 \pm 0.05$ & $\alpha_E^{\rm p,n}=5.60 \pm 0.10$ & \\
& $V^{(3)}_2$ & $\alpha_E^{\rm p,n}=4.54 \pm 0.25$ & $\alpha_E^{\rm p,n}=4.50 \pm 0.28$ & \\
& $V^{(3)}_3$ & $\alpha_E^{\rm p,n}=4.27 \pm 0.23$ & $\alpha_E^{\rm p,n}=4.24 \pm 0.27$ & \\
& $V^{(3)}_4$ & $\alpha_E^{\rm p,n}=3.74 \pm 0.09$ & $\alpha_E^{\rm p,n}=3.69 \pm 0.14$ & \\
& $V^{(3)}_5$ & $\alpha_E^{\rm p,n}=52.6 \pm 5.50$ & $\alpha_E^{\rm p,n}=52.9 \pm 6.10$ & \\
\label{tablealphaEx}
\end{tabular}
\end{table}

\subsubsection{results of Two-body + three-body force model}

In the following we present results of two sets of potentials (cfr. section \ref{twothreeV}): 
i) $V_I$: includes two-body interaction only (see Eq.(\ref{VqqBC})); $V_{III}$: includes a 
three-body term also as in Eq.~(\ref{Vqqqca}). The numerics is summarized in tables
\ref{tableSRqqq} - \ref{tablealphaEqqq}, where also the effects of the hyperfine interaction 
and three-body terms are emphasized. In particular the role of three-body interaction is quite 
important. Neglecting that contribution to $m_3(D'_z)$ yields to a {\it negative} value of the
lower bound (cfr. table \ref{tableLBqqq}), while the hyperfine interaction plays a role similar 
to that described in relation with the hyperradial models. The overall impression is that, 
differently from the hyperradial models, the two body models have a quite small radius 
and therefore a smaller electric polarizability. Adding the three-body term does not improve
the situation and the resulting polarizability is even smaller.

\begin{table}

\caption{Electric dipole sum rule values for the potential models $V_I$, $V_{III}$
(\ref{VqqBC})-(\ref{Vqqqca}). For the $m_3$ sum the contributions arising
from the various potential terms of Eq.~(\ref{m3Dcanoqq}) are shown separately; 
the last value refers to the three-body term and it vanishes for 
the $V_I$ version.}

\begin{tabular}{ccccccccccccccccccccccccccccccccccccccccccccccccccccccccccccccccccccccccccccc}

& ${\rm potentials}$ & $m_0(D'_z)$ & $m_1(D'_z)$ & $m_2(D'_)$ &
$m_3(D'_z)|_{\rm tot}$ & \\
& & [fm$^2$] & [fm] & [u] & [fm$^{-1}$] & \\
\hline
&$V_{I}  $ & $0.0005$ & $0.0014$ & $0.0044$ & $0.0168=0.0022+0.0073+0.0072 + zero$ & \\
&$V_{III}$ & $0.0003$ & $0.0015$ & $0.0197$ & $\;\;\;0.7441=0.1214+0.0040+0.0524+0.5662$ & \\
\label{tableSRqqq}
\end{tabular}
\end{table}

\begin{table}

\caption{Lower bounds to the electric polarizability as predicted by the potential models
$V_I$, $V_{III}$ (\ref{VqqBC})-(\ref{Vqqqca}). The results in brakets () for the 
version $I$ are obtained neglecting the Hyperfine contributions; those ones in [] for the 
version $III$ are obtained excluding the three-body contributions. The polarizability is 
expressed in $10^{-4}$ fm$^3$.}

\begin{tabular}{ccccccccccccccccccccccccccccccccccccccccccccccccccccccccccccccccccccccccccccc}

& ${\rm potentials}$ & $2\,{m_1^2(D'_z)\over m_3(D'_z)}$ &
$2\,{m_0^2(D'_z)\over m_1(D'_z)}$ & 
$2\,{m_0^2(D'_z)\over m_1(D'_z)(1-\Delta/\Gamma)}$ & \\
\hline
& $V_{I}  $ & $2.42$ ($4.25$) & $3.27$ & $3.30$ ($3.24$) & \\
& $V_{III}$ & $0.06$ [$0.25$] & $0.97$ & $\;\;\;1.15$ [$-0.061$] & \\
\label{tableLBqqq}
\end{tabular}
\end{table}

Another interesting conclusion can be drawn for the approximated expression
(\ref{trial3SR}). Its values are still not far from the exact results, but the inclusion 
of three-boy forces worses the situation. One has to emphasize, however, that the bound
$2\,m_0^2/m_1$ cannot be expressed in terms of the mean charge radius of the proton because
of the $SU(6)$ breaking components present in the wave function (\ref{wfcano}) and the 
second identity (\ref{sbound}) is no longer valid. If one assumes $\alpha_E^{\rm p,n} =
{2\over 9} e^2 \left(3 m\right) \langle r^2_{\rm p}\rangle_{\rm ch}$ one should have 
$\alpha_E = 4.37\,10^{-4}$ fm$^3$ for $V_I$ and 
$\alpha_E = 1.12\,10^{-4}$ fm$^3$ for $V_{III}$, values rather far from the exact result
of table \ref{tablealphaEqqq}: the $SU(6)$ breaking components produce relevant effects 
on the polarizability which are not included in the most simple estimates. In particular 
the bounds which involves higher sum rules ($m_3$ and $m_2$) are quite 
sensitive to the tuning effects due to the $\Psi_3$ components and if one puts 
$\Psi_3=0$ in the calculations one cannot satisfy the correct inequalities 
obtaining, for the $V_I$ potential model, a lower bound 
($\alpha_E=3.82\,10^{-3}$ fm$^3$) larger than the upper bound 
($\alpha_E=3.73\,10^{-3}$ fm$^3$).

\begin{table}


\newpage

\caption{Upper bounds to the electric polarizability as predicted by the potential models
$V_I$, $V_{III}$ (\ref{VqqBC})-(\ref{Vqqqca}). The polarizability is expressed in 
$10^{-4}$ fm$^3$.}

\begin{tabular}{ccccccccccccccccccccccccccccccccccccccccccccccccccccccccccccccccccccccccccccc}

& ${\rm potentials}$ & $2\,{m_0(D'_z)\over E^D_{10}}$ & $2\,{m_0(D'_z)\over 
E^D_{10}}\,\Lambda$ & \\
\hline
& $V_{I}  $ & $3.58$ & $3.38$ & \\
& $V_{III}$ & $1.87$ & $1.64$ & \\
\label{tableUBqqq}
\end{tabular}
\end{table}

\begin{table}

\caption{Upper and lower bounds to the nucleon polarizability as predicted by the 
potential models $V_I$, $V_{III}$ (\ref{VqqBC})-(\ref{Vqqqca}).  Notations for the brakets
as in table \ref{tableLBqqq}. The polarizability is expressed in $10^{-4}$ fm$^3$.}

\begin{tabular}{ccccccccccccccccccccccccccccccccccccccccccccccccccccccccccccccccccccccccccccc}

& potentials & $$ & $$ & \\
\hline
& $V_{I}  $ & $\alpha_E^{\rm p,n}=3.34 \pm 0.04$ & ($\alpha_E^{\rm p,n}=3.31 \pm 0.07$) & \\
& $V_{III}$ & $\alpha_E^{\rm p,n}=1.39 \pm 0.24$ & [$\alpha_E^{\rm p,n}=1.42 \pm 0.45$] & \\
\label{tablealphaEqqq}
\end{tabular}
\end{table}

\subsection{Magnetic susceptibility}
\label{masus}

Paramagnetic susceptibility of hyperradial models does not differ from the 
results of Eqs.~(\ref{bNDmodel}) and (\ref{ulb-1}) because the Hamiltonian
(\ref{hyperHo}) commutes with $\mu_z$. On the contrary the $SU(6)$ breaking model
with three-body forces has a much more complex spin structure which is 
evident in Eqs.~(\ref{m0Mqqq}), and (\ref{m1Mqqq}). The results are shown in 
tables \ref{tableMSRqqq} - \ref{tablebetaMqqq}.

\begin{table}

\caption{Values of the magnetic sum rules for the potential models
$V_I$, $V_{III}$ (\ref{VqqBC})-(\ref{Vqqqca}).  }

\begin{tabular}{ccccccccccccccccccccccccccccccccccccccccccccccccccccccccccccccccccccccccccccc}

& ${\rm potentials}$ & $m_0(\mu_z)$ & $m_1(\mu_z)$ & \\
& & [fm$^2$] & [fm] & \\
\hline
&$V_{I}  $ & $5.60\,10^{-4}$ & $0.0011$ & \\
&$V_{III}$ & $6.17\,10^{-4}$ & $0.0015$ & \\
\label{tableMSRqqq}
\end{tabular}
\end{table}

\begin{table}

\caption{Lower bounds to the paramagnetic susceptibility as predicted by
the potential models $V_I$, $V_{III}$ (\ref{VqqBC})-(\ref{Vqqqca}). The 
susceptibility is expressed in $10^{-4}$ fm$^3$.}

\begin{tabular}{ccccccccccccccccccccccccccccccccccccccccccccccccccccccccccccccccccccccccccccc}

& ${\rm potentials}$ & $2\,{m_0^2(\mu_z)\over m_1(\mu_z)}$ & \\
\hline
& $V_{I}  $ & $5.89$ & \\
& $V_{III}$ & $5.23$ & \\
\label{tableMLBqqq}
\end{tabular}
\end{table}

Since we are calculating the simplest bounds only, the uncertainties are larger than
those ones for the polarizability ($13\%$ and $23\%$ for $V_I$ and $V_{III}$ 
respectively), however the reduction with respect the $N$-$\Delta$ approximation 
is quite relevant (roughly $-30\%$) and is due to inclusion of the $\gamma = 2$
component in the hyperspherical expansion (\ref{wfcano}).  
Again the results are identical for protons and neutrons.

\begin{table}

\caption{Upper bounds to the paramagnetic susceptibility as predicted by
the potential models $V_I$, $V_{III}$ (\ref{VqqBC})-(\ref{Vqqqca}). The 
susceptibility is expressed in $10^{-4}$ fm$^3$.}

\begin{tabular}{ccccccccccccccccccccccccccccccccccccccccccccccccccccccccccccccccccccccccccccc}

& potentials & $2\,{m_0(\mu_z)\over E^M_{10}}$ & \\
\hline
& $V_{I}  $ & $7.55$ & \\
& $V_{III}$ & $8.31$ & \\
\label{tableMUBqqq}
\end{tabular}
\end{table}

\begin{table}

\caption{Upper and lower bounds to the paramagnetic susceptibility as predicted by
the potential models $V_I$, $V_{III}$ (\ref{VqqBC})-(\ref{Vqqqca}). The 
susceptibility is expressed in $10^{-4}$ fm$^3$.}

\begin{tabular}{ccccccccccccccccccccccccccccccccccccccccccccccccccccccccccccccccccccccccccccc}

& potentials & $$ & $$ & \\
\hline
& $V_{I}  $ & $\beta_M^{\rm p,n}=6.72 \pm 0.83$ & \\
& $V_{III}$ & $\beta_M^{\rm p,n}=6.77 \pm 1.55$ & \\
\label{tablebetaMqqq}
\end{tabular}
\end{table}

\subsection{Comparison with experiments and conclusions}

In tables \ref{abetaallP}, \ref{abetaallN} all the results are summarized and the retardation
corrections included for both protons and neutrons. The calculation of the additional terms,
within the quark models we are discussing, 
is rather straightforward because they are basically related to the 
electric dipole sums already calculated in the previous sections and 
to the charge proton and neutron radii. In particular (cfr. Eqs.~(\ref{alphae})) 
and (\ref{betam}))
\begin{equation}
\Delta \alpha^{\rm p,n} = {e^2\over 3 M}\,\langle r_{\rm p,n}^2\rangle_{\rm ch}\,\,,
\label{dalphapn}
\end{equation}
\begin{equation}
\Delta \beta^{\rm p,n} = - {3\over 2 M}\,m_0(D'_z) = -{e^2\over 2 M}\,\left(
\langle r^2_{\rm p}\rangle_{\rm ch} + 2\,\langle r^2_{\rm n}\rangle_{\rm ch}
\right)\,\,,
\label{dbetapn}
\end{equation}
\begin{equation}
\beta|_{\rm dia}^{\rm p} =- {e^2\over 6 m}\,\left[\langle r_{\rm p}^2\rangle_{\rm ch}
+{2\over 3}\, \langle r_{\rm n}^2\rangle_{\rm ch}\right]\,,
\label{betadiap}
\end{equation}
\begin{equation}
\beta|_{\rm dia}^{\rm p} =- {e^2\over 6 m}\,\left[\langle r_{\rm n}^2\rangle_{\rm ch}
+{2\over 3}\, \langle r_{\rm p}^2\rangle_{\rm ch}\right]\,,
\label{betadian}
\end{equation}
where also the $SU(6)$ breaking effects have been included. 

Let us discuss first the electric polarizability.

\vspace{2mm}

\noindent The values closest to the experimental data are obtained for the 
hyperradial potential models, where the three-body contributions are included
in the structure of the interaction. The peculiar potential $V_5^{(3)}$ introduced
because it reproduces the electric form factor of the proton, shows the classical
limitation of the quark models: a huge polarizability when one ask to the quark
degrees of freedom to cover all the charge spatial distribution replacing also
the effects of the meson cloud. This is a well known problem and it can be easily 
understood looking at the approximate relation (\ref{aEvar}): by replacing 
the charge root mean square radius with its experimental value one gets
$\alpha_E \approx 42\,10^{-4}$ fm$^3$. On the contrary if one wants to reproduce the
excitation energy of the dipole states $\omega_{\rm h.o.} \approx 600$ MeV,
and one obtains $\alpha_E \approx 3.3\,10^{-4}$ fm$^3$! In order to overcome this
contraddiction we want to investigate better  the predictions of the quark models
which are able to reproduce the spectrum of baryons and look at the quantitative
amount of meson contribution one needs to fill the gap between quark model
contribution and the experimental data.

\begin{table}

\caption{Retardation corrections to the electric polarizability and magnetic susceptibility 
of the proton calculated for the hyperradial potential models (\ref{V1}) - (\ref{V5}) and the
models (\ref{VqqBC})-(\ref{Vqqqca}). The units are $10^{-4}$ fm$^3$.}

\begin{tabular}{ccccccccccccccccccccccccccccccccccccccccccccccccccccccccccccccccccccccccccccc}

& ${\rm potentials}$ & $\Delta \alpha_E^{\rm p}$ & $\bar \alpha_E^{\rm p}$ & 
$\Delta \beta^{\rm p}$ & $\beta|^{\rm p}_{\rm dia}$ & $\bar \beta_M^{\rm p}$ & \\
\hline
&$V^{(3)}_1$ & $1.36$ & $6.96$ & $-2.04$ & $-2.04$ & $4.61$ & \\
&$V^{(3)}_2$ & $1.19$ & $5.70$ & $-1.79$ & $-1.79$ & $5.13$ & \\
&$V^{(3)}_3$ & $1.16$ & $5.40$ & $-1.73$ & $-1.73$ & $5.23$ & \\
&$V^{(3)}_4$ & $1.09$ & $4.78$ & $-1.64$ & $-1.64$ & $5.24$ & \\
&$V^{(3)}_5$ & $3.93$ & $56.8$ & $-5.90$ & $-5.90$ & $-3.10$ & \\
\hline
&$V_I$       & $1.22$ & $4.56$ & $-1.41$ & $-1.60$ & $3.71$ & \\
&$V_{III}$   & $0.61$ & $2.00$ & $-0.83$ & $-0.87$ & $5.06$ & \\
\label{abetaallP}
\end{tabular}
\end{table}

In the case of hyperradial TBM potentials let us take the results coming from the
linear plus confining  "Coulomb" (i.e. the interaction $V_1^{(3)}$) as the typical
example. Which contribution should one need from mesons to reproduce the experimetal 
polarizabilities of neutron and proton? Including the meson contributions into
the retardation corrections (\ref{dalphapn}) is rather simple by taking the experimental 
value of the proton charge radius: one obtains the prediction for the dynamic 
polarizability $\bar \alpha_E^{\rm p} = (5.60 + 3.78)\,10^{-4} = 9.38\,10^{-4}$ fm$^3$
which is claming for a contribution of the mesons in the range of $\approx 2.7\,10^{-4}$
fm$^3$. For the neutron the static value of the polarizability is known experimentally
(cfr. section \ref{intro}) and the meson contribution is found to be much larger
$\approx 7.3\,10^{-4}$ fm$^3$ in agreement with the calculation of ref.\cite{soli2}
where the difference between the neutron and proton static polarizability is found
$5\,10^{-4}$ fm$^3$ and ascribed to mesonic degrees of freedom. The net result is that 
the contribution of the quarks to the proton polarizability is not small and comparable with 
the one coming from the meson cloud. This conclusion is basically valid for all the TBM 
$V_1^{(3)}-V_3^{(3)}$ fitted on the excitation spectrum and it favours the QCD "inspired" 
form $-\tau/\xi + k_l\,\xi$, i.e. the $V_1^{(3)}$ potential model. On the contrary the 
potential models where the three-body forces are included phenomenologically to reproduce 
the position of the Roper resonance, predict quite low values of the static polarizability
therefore asking for a much larger meson contribution \cite{lvov93}.

\vspace{2mm}

\noindent The results on the magnetic susceptibilities 

\vspace{2mm}

\noindent The predicted paramagnetic susceptibility is the same for all 
the TBM potentials $\beta_M^{\rm p,n}|_{\rm para}=8.7\,10^{-4}$ fm$^3$. In 
particular for the proton the inclusion of meson spatial distribution on the 
correction (\ref{dbetapn}) and on the diamagnetic contribution (\ref{betadiap})
yields to $\bar \beta_M^{\rm p} = (8.7 - 9.0 )\,10^{-4}$ fm$^3$ and therefore
to a mesonic contribution $2.4 \pm 1.3\,10^{-4}$ fm$^3$, a result consistent 
with the analysis of ref.\cite{muko93} where the $M1$ pion photoproduction
has been investigated.

\begin{table}

\caption{Retardation corrections to the electric polarizability and magnetic susceptibility 
of the neutron calculated for the hyperradial potential models (\ref{V1}) - (\ref{V5}) and the
models (\ref{VqqBC})-(\ref{Vqqqca}). The units are $10^{-4}$ fm$^3$.}

\begin{tabular}{ccccccccccccccccccccccccccccccccccccccccccccccccccccccccccccccccccccccccccccc}

& ${\rm potentials}$ & $\Delta \alpha_E^{\rm n}$ & $\bar \alpha_E^{\rm n}$ & 
$\Delta \beta^{\rm n}$ & $\beta|^{\rm n}_{\rm dia}$ & $\bar \beta_M^{\rm n}$ & \\
\hline
&$V^{(3)}_1$ & $0$ & $5.60$ & $-2.04$ & $-1.36$ & $5.29$ & \\
&$V^{(3)}_2$ & $0$ & $4.50$ & $-1.78$ & $-1.19$ & $5.72$ & \\
&$V^{(3)}_3$ & $0$ & $4.24$ & $-1.73$ & $-1.16$ & $5.81$ & \\
&$V^{(3)}_4$ & $0$ & $3.69$ & $.1.64$ & $-1.09$ & $5.97$ & \\
&$V^{(3)}_5$ & $0$ & $52.9$ & $-5.90$ & $-3.93$ & $-1.16$ & \\
\hline
&$V_I$       & $-0.10$ & $3.24$ & $-1.41$ & $-0.99$ & $4.32$ & \\
&$V_{III}$   & $-0.02$ & $1.37$ & $-0.83$ & $-0.57$ & $5.37$ & \\
\label{abetaallN}
\end{tabular}
\end{table}

The neutron susceptibility has been experimentally derived by the knowledge
of the static electric polarizability, adding retardation corrections and using 
the dispersion relation (\ref{baldsr}), the precision is rather limited. 
The mesonic contribution can be extracted from 
$\bar \beta_M^{\rm n} = (8.7 - 6.8 )\,10^{-4}$ fm$^3$ 
and it results to be $1.9 \pm 3.6\,10^{-4}$ fm$^3$.

The potential models $V_I$ - $V_{III}$ give analogous results. 
The remarkable difference is that their complicate spin structure can be 
easily seen looking at
the mean excitation energy $m1(\mu_z)/m_0(\mu_z) \approx 370$ MeV and $467$ MeV
respectively which is larger than the $M_\Delta - M_N$ indicating that the
$SU(6)$-breaking mechanism introduces a larger excitation spectrum: an effect
which results also in the width of the predictions for $\beta_M$ in table 
\ref{tablebetaMqqq}.


\begin{thebibliography}{9}
\bibitem{notehyd}		A simple example is given by the (atomic) 
				hydrogen.
 The exact value of the electric 
				polarizability is $\alpha_E = 9/2\,a_0^3$
				(where $a_0$ is the Bohr's radius). 
				Considering only the first two excited 
				states one
				would get $\alpha_E = (2.96 + 0.40)\,a_0^3$, 
				i.e. $75\%$ of the exact result.
\bibitem{note1}		The Eq.~(\ref{m+1D}), valid for the whole class of velocity indipendent 
			potentials, is similar to the analogous
{\sl atomic}
			sum rule for electric dipole excitations rather than the corresponding
			{\sl nuclear} sum where an enhancement factor due to meson-exchange 				
			contributes to the total sum. In fact gluons (similarly to photons)
			do not carry electric charges and "exchange currents" present in 
			the nucleon-nucleon interaction vanish in the present case as 
			well as for electrons in the atoms.
\bibitem{notevar}		Variational approaches, which are restricted to a 
				subspace of all the possible solutions, give an upper 
				bound to the total energy of the system and, consequentely, a 
				lower bound to the polarizability through the relation
				$E_{\rm tot}= E_0 - {1\over 2}\,\alpha_E\,{\cal E}^2$.
\bibitem{note3}			An exception is the harmonic oscillator potential which has a 
				pure two-body character, but can be expressed by means of the
				hyperradius $\xi$ in the following way:
				\begin{equation}
				V_{\rm h.o.}={1\over 2}\,K\,\sum_{i<j}({\bf r}_i-{\bf r}_j)^2 
				= {3\over 2}\,K \left({\vec \rho}\,^2+{\vec \lambda}^2\right) 
				= {3\over 2}\,K\,\xi^2\,\,.
				\label{hyperhosc}
				\end{equation}
\bibitem{note4}			The discrepancy involves more than the nucleon resonances. In fact 
				it concerns the first excited state with the same quantum numbers as 
				the ground state, namely, the Roper $N(1440)$, and the analogous 
				levels $\Delta(1600)$, $\Lambda(1600)$ and $\Sigma(1660)$.
\bibitem{notedelta} The second equality of Eq.~(\ref{m3xHyp}) is valid as long as 
                    the integral $\int d\xi\,\xi\,\psi_{00}(\xi)\,\psi'_{00}(\xi)$ 
                    coverges, in the other cases one has to resort to a numerical 
                    evaluation of the laplacian of the delta function as a limit of 
                    some smooth version, like the gaussian form $\delta({\vec \rho\,}) 
                    \sim \exp(-{\vec \rho\,}^2/\epsilon^2)$ with $\epsilon \to 0$. 
                    In this case the Eq.~(\ref{m3xHyp})becomes:
\begin{equation}
\left. m_3(D'_z)\right|_{\rm Hyp} = \lim_{\epsilon \to 0}\,
\left[-{4\over 9}\,{e^2\over m^2}\,{M_\Delta-M_N \over \epsilon^7 \pi \sqrt \pi}\,
{{\cal I}(\epsilon^2)\over \int d\xi\,\xi^2\,|\psi_{00}(\xi)|^2}\right]\,\,,\\
\nonumber
\end{equation}
with
\begin{equation}
{\cal I}(\epsilon^2) = \int_0^1 dt\,\int d\xi\,t^2\,\sqrt{1-t^2}\,\left(2\,\xi^2\,t^2
-3\,\epsilon^2\right)\,\xi^5\,{\rm e}^{-\xi^2\,t^2/\epsilon^2}\,|\psi_{00}(\xi)|^2
\,\,,\\
\nonumber
\end{equation}
                     and the convergence has to be checked numerically.
\end{thebibliography}
\end{document}